\documentclass{rmf-d}
\usepackage{nopageno,rmfbib,multicol,times,epsf,amsmath,amssymb,cite}
\usepackage[latin1]{inputenc}
\usepackage[]{caption2}
\usepackage{graphicx}
\usepackage{url}
\usepackage{hyperref}

\def\ba{\begin{eqnarray}}
\def\ea{\end{eqnarray}}
\def\be{\begin{equation}}
\def\ee{\end{equation}}

\def\rmfcornisa{}

\begin{document}
\title{   Bottomonium suppression and elliptic flow in heavy-ion collisions
\vspace{-6pt}}
\author{ Michael Strickland }
\address{ Department of Physics, Kent State University, Kent, OH 44242 United States    }
\maketitle
%\recibido{day month year}{day month year \vspace{-12pt}}
\begin{abstract}
\vspace{1em} 
In this proceedings contribution I review recent progress concerning the suppression of bottomonium production in the quark-gluon plasma.  Making use of open quantum system methods applied to potential non-relativistic quantum chromodynamics one can show that the dynamics of heavy-quarkonium bound states satisfying the scale hierarchy $1/a_0 \gg \pi T \sim m_D \gg E$ obey a Lindblad equation whose solution provides the quantum evolution of the heavy-quarkonium reduced density matrix.  To solve the resulting Lindblad equation we use a quantum trajectories algorithm which allows one to include all possible angular momentum states of the quark-antiquark probe in a scalable manner.  We solve the Lindblad equation using a tuned 3+1D dissipative hydrodynamics code for the background temperature evolution.  We then consider a large number of Monte-Carlo sampled bottomonium trajectories embedded in this background.  This allows us to extract the centrality- and $p_T$-dependence of the nuclear suppression factor $R_{AA}[\Upsilon]$ and elliptic flow $v_2[\Upsilon]$.  We find good agreement between our model predictions and available $\sqrt{s_{NN}}$ = 5.02 TeV Pb-Pb collision experimental data from the ALICE, ATLAS, and CMS collaborations.
\vspace{1em}
\end{abstract}
\keys{  Bottomonium suppression, Effective field theory methods, Open quantum system methods, Quark-gluon plasma, Relativistic heavy-ion collisions, Quantum chromodynamics  \vspace{-4pt}}
\pacs{   12.38.Mh, 14.40.Pq, 14.40.Nd,  25.75.-q    \vspace{-4pt}}
\begin{multicols}{2}

\section{Introduction}

The production of heavy-quark bound states in nucleus-nucleus collisions relative to nucleon-nucleon collisions can provide much needed information about the space-time evolution of the quark-gluon plasma (QGP), including the effective temperature of the plasma created and, in the case of harmonic flows, constraints on the collision geometry and flow.  Historically, it was first suggested by Matsui and Satz \cite{Matsui:1986dk} that heavy quarkonium production would be suppressed in high-energy heavy-ion collisions based on the fact that Debye screening in the plasma would reduce the effective interaction between quarks and anti-quarks, eventually leading to the disassociation of all color singlet bound states.

The screening model of disassociation was, however, missing a key ingredient, namely that, even below the disassociation temperature, it is possible for states to become unbound through in-medium gluon exchange.  Such in-medium breakup is encoded in the in-medium width of singlet states which, in the heavy-quark limit, emerges due to the imaginary part of the heavy-quark potential.  In a finite temperature QGP there is a QED-like effect resulting from Landau damping of the exchanged gluon \cite{Laine:2006ns,Dumitru:2009fy} and also a genuinely non-abelian effect resulting from transitions between color singlet and octet states \cite{Brambilla:2008cx,Beraudo:2007ky,Escobedo:2008sy,Brambilla:2010vq,Brambilla:2011sg,Brambilla:2013dpa}.  Applications of realistic complex-valued potential models to QGP phenomenology have demonstrated good agreement with experimental measurements of bottomonium suppression~\cite{Strickland:2011mw,Strickland:2011aa,Krouppa:2015yoa,Krouppa:2017jlg,Islam:2020bnp,Islam:2020gdv}, however, these studies have not included non-potential effects associated with quantum jumps.  

In recent years a new formulation of heavy-quarkonium suppression in the QGP has been developed which allows one to include the effect of in-medium color and angular momentum transitions.  This formulation relies on the application of open quantum system (OQS) methods to obtain {\em master equations} which can describe the time evolution of the heavy-quarkonium reduced density matrix \cite{Beraudo:2007ky,Brambilla:2008cx,Escobedo:2008sy,Rothkopf:2011db,Akamatsu:2011se,Akamatsu:2014qsa,Blaizot:2015hya,Blaizot:2017ypk,Yao:2018nmy,Blaizot:2018oev,Yao:2020xzw,Rothkopf:2019ipj,Miura:2019ssi,Yao:2020eqy,Akamatsu:2020ypb,Brambilla:2020qwo,Brambilla:2021wkt,Omar:2021kra}.  Although not guaranteed in general, in many cases these master equations can be cast into Lindblad form \cite{Gorini:1975nb,Lindblad:1975ef} which makes them easier to solve numerically.   In this proceedings contribution we will review recent work in which OQS methods are applied within the \mbox{pNRQCD} effective field theory \cite{Pineda:1997bj,Brambilla:1999xf,Brambilla:2004jw}. In the case of a strongly-coupled QGP with the scale hierarchy $1/a_0 \gg \pi T \sim m_D \gg E$, the pNRQCD master equation can be cast in Lindblad form, where $a_0$ is the Bohr radius, $T$ is the temperature, $m_D$ is the Debye screening mass, and $E$ is the binding energy of the state.  Due to this, it is possible to solve the evolution equation using a method called the {\em quantum trajectories algorithm} \cite{Dalibard:1992zz,Omar:2021kra} which can be deployed in a massively parallel manner.  This allows one to solve the Lindblad equation without cutoffs in the angular momentum and to straightforwardly compute results for a large set of sampled physical paths through the QGP.  Using this method we make predictions for the centrality and $p_T$ dependence of $R_{AA}[\Upsilon]$ and $v_2[\Upsilon]$.  

%%%%%%%%%%%%%%%%%%%%%%%%%%%%%%%%%%%%%%%%%%%%%%%%%%%%%%%%%%%
\begin{figure*}[t!]
\centerline{
\includegraphics[width=0.465\linewidth]{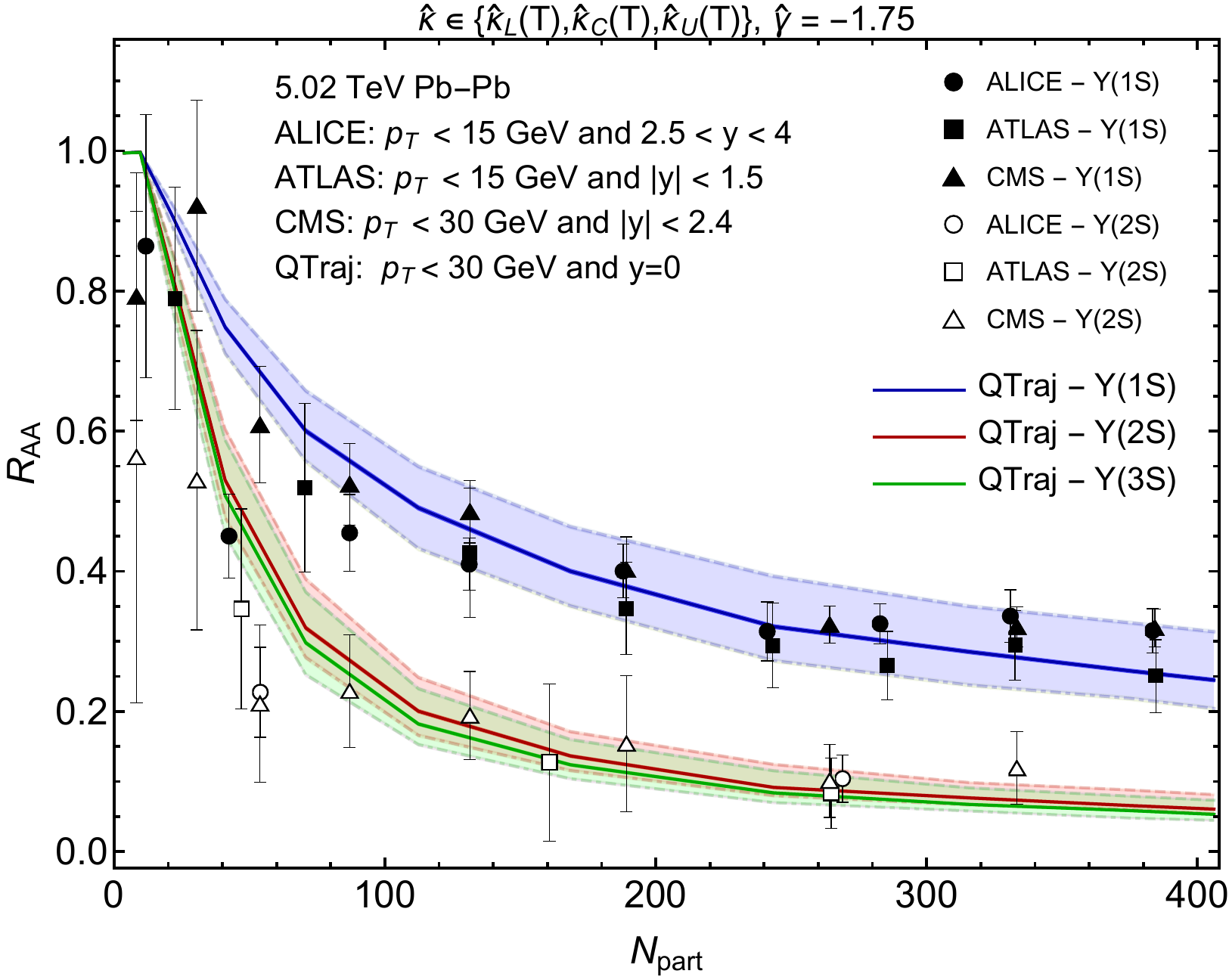}\hspace{2mm}
\includegraphics[width=0.465\linewidth]{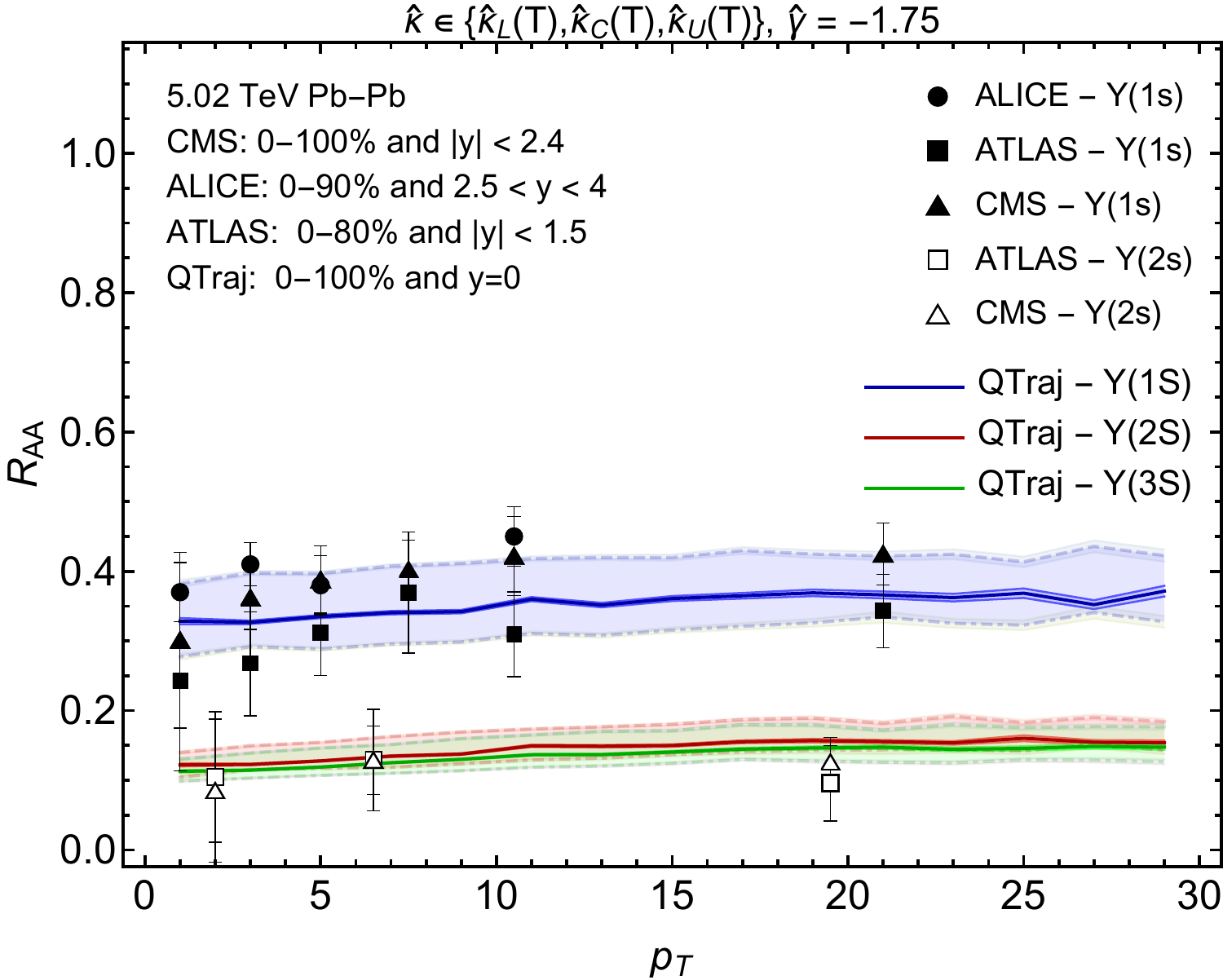}
}
\vspace{1mm}
\caption{(Color online)
The nuclear modification factor of $\Upsilon(1S)$, $\Upsilon(2S)$, and $\Upsilon(3S)$ states as a function of $N_{\text{part}}$.
The bands indicate variation with respect to $\hat{\kappa}(T)$ (left) and $\hat{\gamma}$ (right).
The central curves represent the central values of $\hat{\kappa}(T)$ and $\hat{\gamma}$.  In both panels we show experimental measurements from the ALICE~\cite{ALICE:2020wwx}, ATLAS~\cite{ATLAS5TeV}, and CMS~\cite{Sirunyan:2018nsz} collaborations.
}
\label{fig:raa}
\end{figure*}
%%%%%%%%%%%%%%%%%%%%%%%%%%%%%%%%%%%%%%%%%%%%%%%%%%%%%%%%%%%

\section{Methods}
\label{sec:methods}

Using OQS methods applied within the pNRQCD effective field theory framework Refs. \cite{Brambilla:2016wgg,Brambilla:2017zei} obtained a master equation for heavy quarkonium evolution in a strongly coupled equilibrium QGP.  When the temperature is much larger than the binding energy, $T \gg E$, one can make an expansion in $E/T$, finding at leading order an evolution equation of Lindblad form~\cite{Brambilla:2016wgg,Brambilla:2017zei,Lindblad:1975ef,Gorini:1975nb}
\begin{equation}\label{eq:lindblad}
\frac{d \rho(t)}{dt} = -i[H, \rho(t)] + \sum_{n} \! \left( \! C_{n} \rho(t) C_{n}^{\dagger} - \frac{1}{2} \! \left\{ C_{n}^{\dagger}C_{n}, \rho(t) \right\} \! \right),
\end{equation}
where, for $N_c$ colors, the reduced density matrix and Hamiltonian are given by
\begin{align}
\rho(t) =& \begin{pmatrix} \rho_{s}(t) & 0 \\ 0 & \rho_{o}(t) \end{pmatrix},\\
H =& \begin{pmatrix} h_{s} & 0 \\ 0 & h_{o} \end{pmatrix} + \frac{r^{2}}{2} \gamma \begin{pmatrix} 1 & 0 \\ 0 & \frac{N_{c}^{2}-2}{2(N_{c}^{2}-1)} \end{pmatrix} ,
\end{align}
with $\rho_{s}(t)$ and $\rho_{o}(t)$ being the singlet and octet reduced density matrices, respectively, and $h_{s,o}={\mathbf{p}^{2}}/{M} + V_{s,o}$ being the singlet or octet Hamiltonian.
The jump (collapse) operators $C$ appearing in Eq.~\eqref{eq:lindblad} are 
\ba
C_{i}^{0} &=& \sqrt{\frac{\kappa}{N_{c}^{2}-1}} r^{i} \begin{pmatrix} 0 & 1 \\ \sqrt{N_{c}^{2}-1} & 0 \end{pmatrix}, \\
C_{i}^{1} &=& \sqrt{\frac{(N_{c}^{2}-4)\kappa}{2(N_{c}^{2}-1)}} r^{i} \begin{pmatrix} 0 & 0 \\ 0 & 1 \end{pmatrix},
\ea
with $i \in \{1,2,3\}$ and the transport coefficients $\kappa$ and $\gamma$ given by the following chromoelectric correlators 
\ba
\kappa &=& \frac{g^{2}}{18} \int_{0}^{\infty} dt \left\langle \left\{ \tilde{E}^{a,i}(t,\mathbf{0}), \tilde{E}^{a,i}(0,\mathbf{0}) \right\} \right\rangle, \\
\gamma &=& -i \frac{g^{2}}{18} \int_{0}^{\infty} dt \left\langle \left[ \tilde{E}^{a,i}(t,\mathbf{0}), \tilde{E}^{a,i}(0,\mathbf{0}) \right] \right\rangle  ,
\ea
which can be measured directly on the lattice.

For the background evolution, we use aHydro3p1 1.0, which is a three-dimensional dissipative anisotropic hydrodynamics code.
The aHydro3p1 code implements quasiparticle anisotropic hydrodynamics and, as a result, has a realistic lattice-based equation of state \cite{Martinez:2010sc,Florkowski:2010cf,Alqahtani:2015qja,Alqahtani:2017mhy}.  The initial central temperature and shear viscosity to entropy density have been tuned to soft observables such as the pion, kaon, and proton spectra and elliptic flow collected at \mbox{$\sqrt{s_{NN}} =$ 5.02 TeV}~\cite{Alqahtani:2020paa}.

\section{Results}
\label{sec:results}

We solve \eqref{eq:lindblad} using a quantum trajectories implementation called QTraj 1.0~\cite{Omar:2021kra}.  The QTraj code is available for free use under the GNU Public License version 3.  Download and installation instructions can be found in Ref.~\cite{Omar:2021kra}.  We use lattice quantum chromodynamics measurements to constrain the transport coefficients $\kappa$ and $\gamma$.  We vary $\kappa$ over three temperature-dependent parameterizations \mbox{$\hat\kappa(T) = \kappa(T)/T^3 \in \{ \hat\kappa_L(T), \hat\kappa_C(T),\hat\kappa_U(T)\}$} with the lower, central, and upper limits taken from Ref.~\cite{Brambilla:2020siz}.  For $\gamma$, in order to encompass all current indirect lattice extractions, we take $\hat{\gamma} = \gamma / T^{3} = \{-3.5,\, -1.75, \, 0 \}$ \cite{Brambilla:2019tpt,Kim:2018yhk,Aarts:2011sm,Larsen:2019bwy,Shi:2021qri}.  At leading order in $E/T$, the coefficients $\hat\kappa$ and $\hat\gamma$ set the magnitude of the imaginary and real parts of the potential, respectively.

To calculate the nuclear suppression in $AA$ collisions, we compute the survival probability for a large number of physical trajectories.  Each physical trajectory has a Monte-Carlo-sampled initial production point and transverse momentum.  Along each physical trajectory we then average over a large set of quantum trajectories in which the quantum evolution is stochastically realized.  This procedure allows one to compute the QGP survival probability for all bound states.  We then perform late-time excited state feed down using a feed down matrix $F$ which is constructed from measured branching ratios and cross sections for bottomonium states~\cite{Brambilla:2020qwo}.  

%%%%%%%%%%%%%%%%%%%%%%%%%%%%%%%%%%%%%%%%%%%%%%%%%%%%
\begin{figure*}[t]
	\begin{center}
		\includegraphics[width=0.46\linewidth]{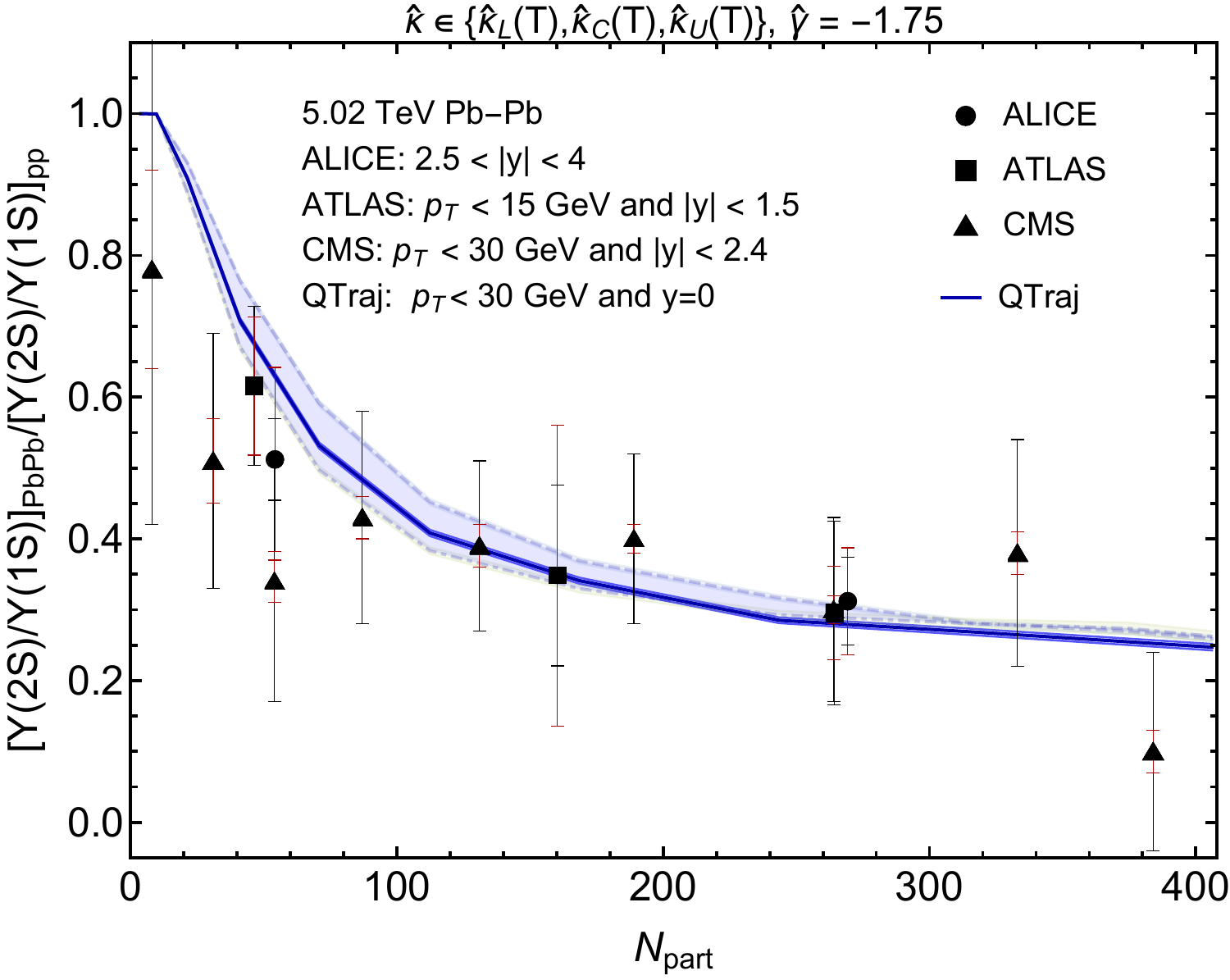} \hspace{4mm}
		\includegraphics[width=0.46\linewidth]{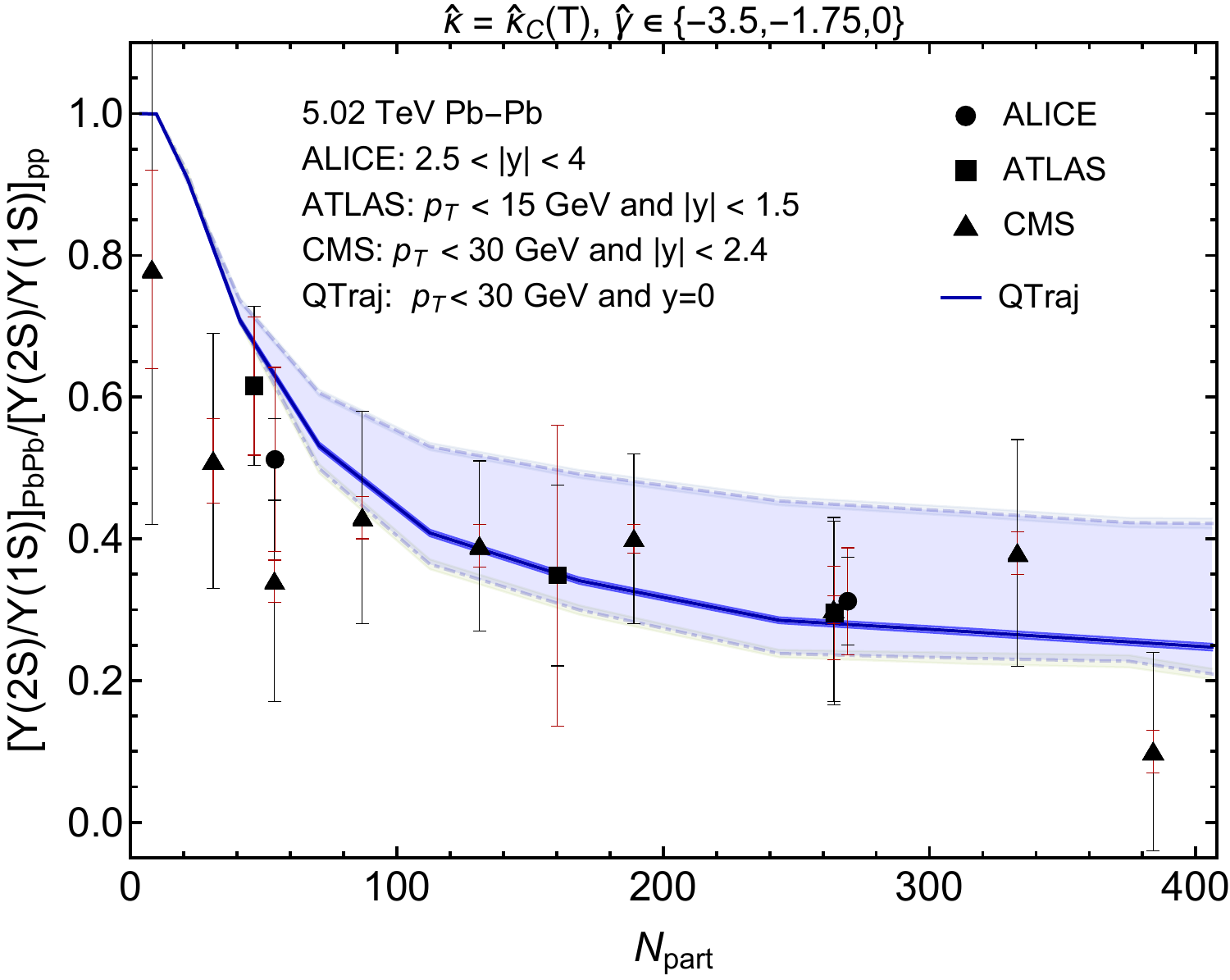}
	\end{center}
	\vspace{-5mm}
	\caption{(Color online)
		The double ratio of the nuclear modification factor $R_{AA}[\Upsilon(2S)]$ to $R_{AA}[\Upsilon(1S)]$ as a function of $N_{\text{part}}$ compared to experimental measurements of the ALICE~\cite{ALICE:2020wwx}, ATLAS~\cite{ATLAS5TeV}, and CMS~\cite{CMS:2017ycw} collaborations.
		The bands indicate variation of $\hat{\kappa}(T)$ and $\hat{\gamma}$ as in Fig.~\ref{fig:raa}.
		The black and red experimental error bars represent statistical and systematic uncertainties, respectively.
	}
	\label{fig:2s_double_ratio_vs_npart}
\end{figure*}
%%%%%%%%%%%%%%%%%%%%%%%%%%%%%%%%%%%%%%%%%%%%%%%%%%%%

%%%%%%%%%%%%%%%%%%%%%%%%%%%%%%%%%%%%%%%%%%%%%%%%%%%%
\begin{figure*}[t]
	\begin{center}
		\includegraphics[width=0.46\linewidth]{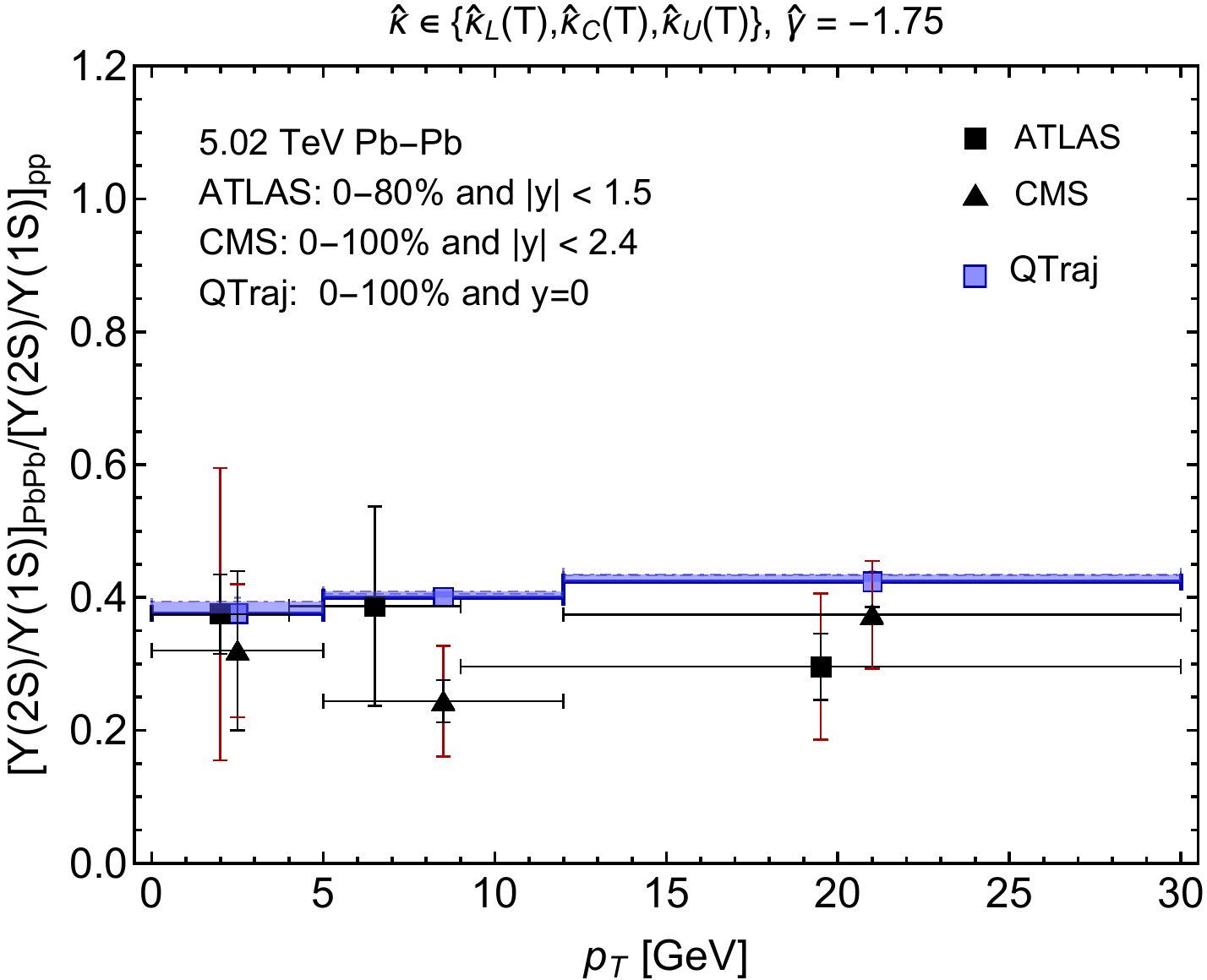}  \hspace{4mm}
		\includegraphics[width=0.46\linewidth]{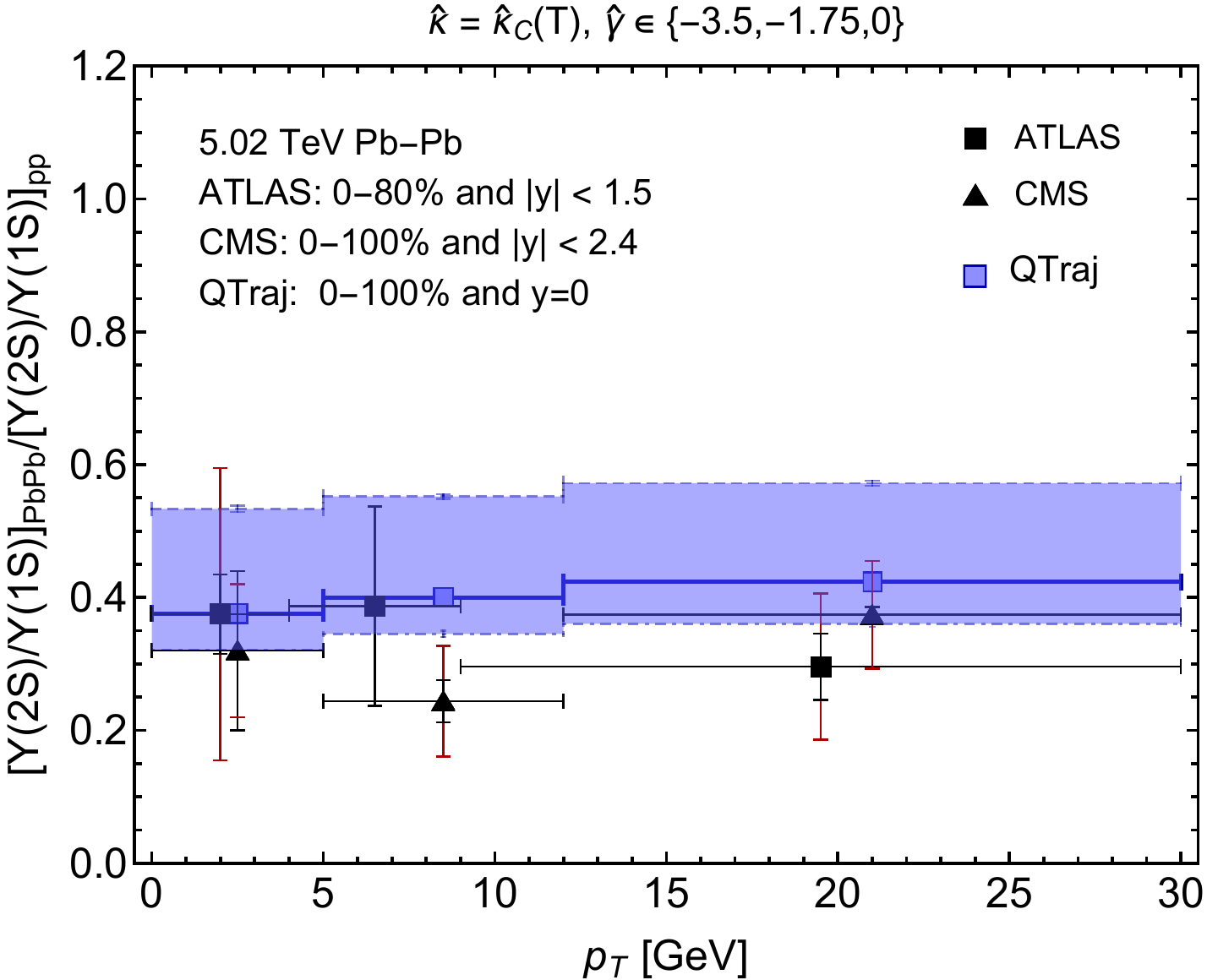}
	\end{center}
	\caption{(Color online)
		The double ratio of the nuclear modification factor $R_{AA}[\Upsilon(2S)]$ to $R_{AA}[\Upsilon(1S)]$ as a function of $p_{T}$ compared to experimental measurements of the ATLAS~\cite{ATLAS5TeV}, and CMS~\cite{CMS:2017ycw} collaborations.
		The bands and bars represent uncertainties as in Fig.~\ref{fig:2s_double_ratio_vs_npart}.
	}
	\label{fig:2s_double_ratio_vs_pt}
\end{figure*}
%%%%%%%%%%%%%%%%%%%%%%%%%%%%%%%%%%%%%%%%%%%%%%%%%%%%

%%%%%%%%%%%%%%%%%%%%%%%%%%%%%%%%%%%%%%%%%%%%%%%%%%%%
\begin{figure*}[t]
	\begin{center}
		\includegraphics[width=0.465\linewidth]{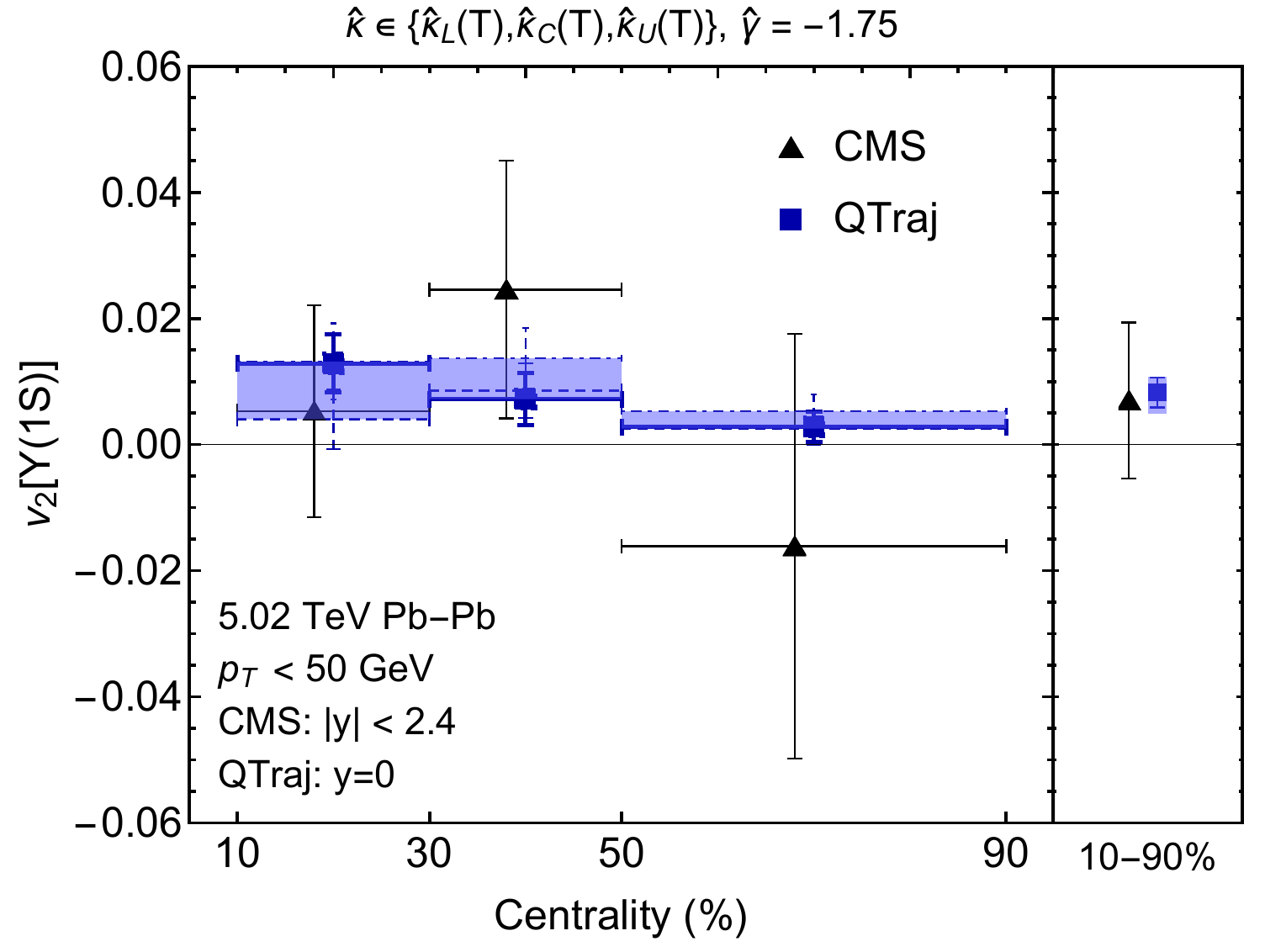} \hspace{2mm}
		\includegraphics[width=0.465\linewidth]{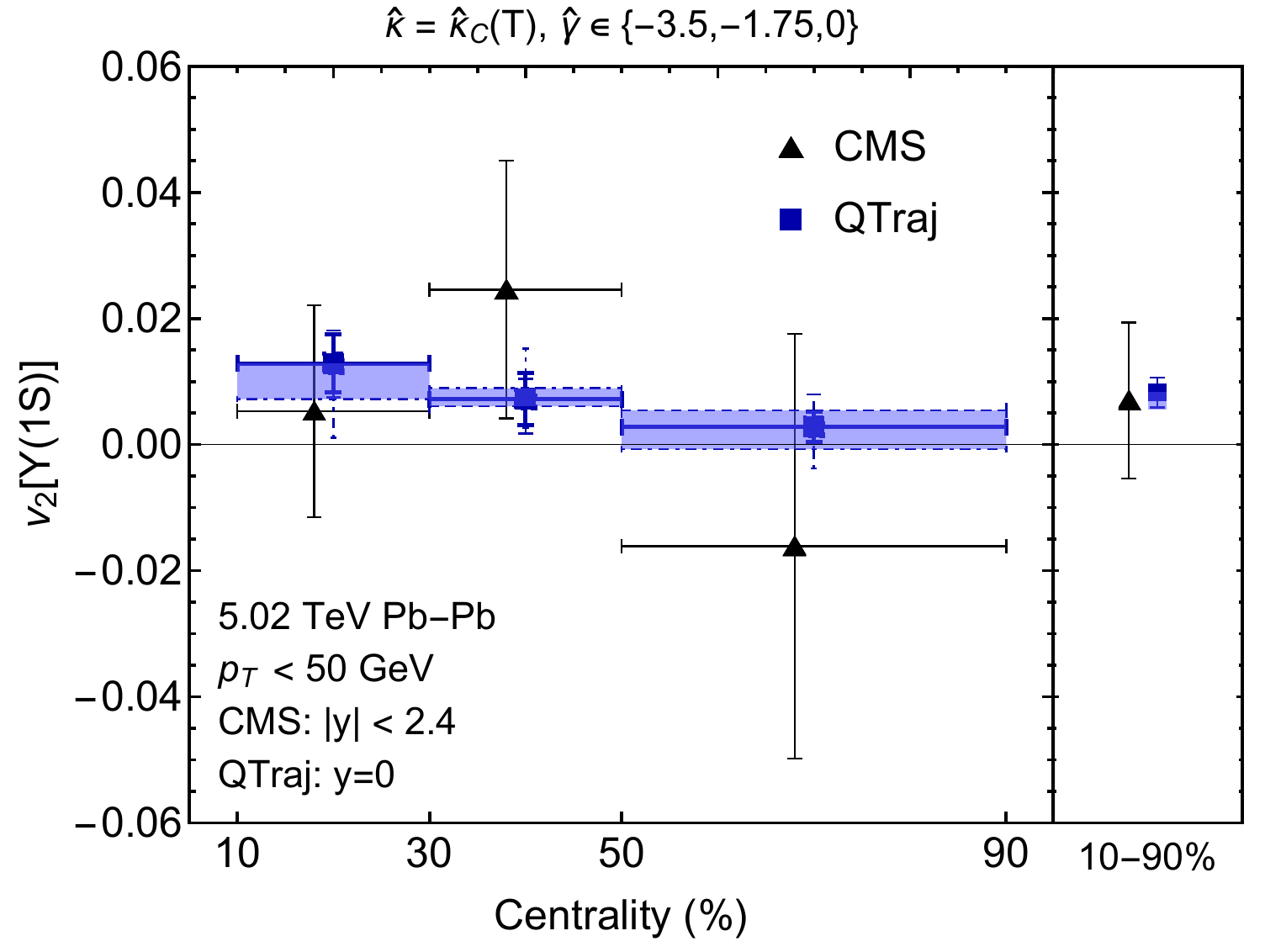}\;\;\;\;
	\end{center}
	\vspace{-5mm}
	\caption{(Color online)
		The elliptic flow $v_{2}$ of the $\Upsilon(1S)$ as a function of centrality compared to experimental measurements of the CMS~\cite{CMS:2020efs} collaboration.
		The QTraj bands represent uncertainties as in Fig.~\ref{fig:raa}.  QTraj error bars indicate the statistical uncertainty of our extraction.
	}
	\label{fig:v2_1S_vs_centrality}
\end{figure*}
%%%%%%%%%%%%%%%%%%%%%%%%%%%%%%%%%%%%%%%%%%%%%%%%%%%%

In practice, we obtain the nuclear suppression of state $i$ using
\be
R^{i}_{AA}(c,p_T,\phi) = \frac{\left(F \cdot S(c,p_T,\phi) \cdot \vec{\sigma}_{\text{direct}}\right)^{i}}{\vec{\sigma}_{\text{exp}}^{i}} \, ,
\label{eq:feeddown}
\ee
where $ \vec{\sigma}_{\text{direct}}$ and $\vec{\sigma}_{\text{exp}}$ are the pre feed down and experimentally observed $pp$ cross sections for bottomonium production, respectively.  The labels $c$, $p_T$, and $\phi$ correspond to the centrality, transverse momentum, and azimuthal angle bins considered.  The survival probability $S(c,p_T,\phi)$ is diagonal, with the entries being each state's survival probability.  

In practice we average over millions of sampled physical trajectories and use a large one-dimensional lattice to solve for the stochastic wave function evolution.  For the precise lattice spacing, time steps, initial conditions used, etc., we refer the reader to Ref.~\cite{Brambilla:2021wkt}.  In the left panel of Fig.~\ref{fig:raa} we plot the nuclear suppression factor as a function of the number of participants.  The band in the figure indicates the theoretical uncertainty in $R_{AA}$ stemming from our uncertainty in the transport coefficient $\kappa$.  Similar results are obtained when varying $\gamma$, however, the band is somewhat larger~\cite{Brambilla:2021wkt}.  As can be seen from  Fig.~\ref{fig:raa}, our predictions for the dependence of $R_{AA}$ on $N_\text{part}$ agree quite well with the reported experimental data.  In the right panel of Fig.~\ref{fig:raa} we present our predictions for the $p_T$-dependence of $R_{AA}$ and compare to experimental data.  Once again we see good agreement given current theoretical and experimental uncertainties.

In Figs.~\ref{fig:2s_double_ratio_vs_npart} and \ref{fig:2s_double_ratio_vs_pt} we present our predictions for the $N_{\rm part}$ and $p_T$ dependence of the 2S to 1S double ratio compared to experimental data from the ATLAS and CMS collaborations.  In the left panels we show the result of varying $\hat\kappa$ and in the right panels we show the result of varying $\hat\gamma$.  As these figures demonstrate, the 2S to 1S double ratio is insensitive to $\hat\kappa$, however, is quite sensitive to $\hat\gamma$.  This pattern is repeated also in the 3S to 1S double ratio as shown in Ref.~\cite{Brambilla:2021wkt}.  As a result, it may be possible to constrain $\hat\gamma$ based on fits to experimental data for these double ratios.

Finally, in Fig.~\ref{fig:v2_1S_vs_centrality} we present predictions for the elliptic flow of the $\Upsilon(1S)$ in three different centrality classes.  We additionally show the centrality-integrated result in the right sub-panel of both figures. The left and right figures in Fig.~\ref{fig:v2_1S_vs_centrality} correspond to varying $\hat\kappa$ and $\hat\gamma$, respectively.  In the case of $v_2[\Upsilon(1S)]$, the variations with $\hat\kappa$ and $\hat\gamma$ are similar and our predictions are, again, in reasonable agreement with available data.  Focussing on the integrated results shown in the right sub-panel of each figure, we find that in this case (a) our predictions have small systematic/statistical uncertainties and (b) are consistent with CMS measurements within experimental uncertainties.

\section{Outlook}

In this proceedings contribution, I presented a collection of predictions for bottomonium suppression in \mbox{$\sqrt{s_{NN}}$ = 5.02 TeV} Pb-Pb collisions.  To make predictions for $R_{AA}[\Upsilon]$ and $v_2[\Upsilon]$ we made use of a publicly released code that solves the Lindblad equation using the quantum trajectories algorithm \cite{Omar:2021kra}.  The physics predictions of this code were summarized here and additional theory/data comparisons plus a more in-depth discussion of our results can be found in Ref.~\cite{Brambilla:2021wkt}.  The work reported here represents the first stochastic OQS approach to bottomonium suppression that includes full 3D quantum evolution, the non-abelian QCD physics of singlet-octet transitions, and a realistic 3+1D hydrodynamical background.  This extends earlier work on bottomonium transport in the QGP~\cite{Grandchamp:2005yw,Emerick:2011xu,Du:2017qkv,Du:2019tjf} to include explicit quantum dynamics and internal transitions.  

We made use of pNRQCD and OQS methods to obtain the Lindblad equation that emerges when one truncates the pNRQCD master equation at leading order in the ratio of the binding energy to temperature, $E/T$.  The resulting Lindblad equation can be solved numerically if one discretizes the wave function and decomposes the states into color and angular momentum eigenstates, however, in practice, one finds that the size of the matrix equation is prohibitive.  To work around this, we made use of an algorithm originally developed for quantum optics applications, called the quantum trajectories algorithm.  In this way, solution of the 3D non-abelian Lindblad equation could be reduced to the solution of a 1D Schr\"odinger equation with a non-Hermitian Hamiltonian subject to stochastic internal transitions (in angular momentum and color) called quantum jumps.  This algorithm allows one to solve the 3D non-abelian Lindblad equation in a massively parallel manner since each physical and sub-sampled quantum trajectory are independent.

One of the limitations of the treatment in Ref.~\cite{Brambilla:2021wkt} is that the master equation was truncated at order $(E/T)^0$.  The first sub-leading correction occurs at order $(E/T)^1$.  Since, in the relevant temperature range, $E/T$ need not be small, it seems necessary to include these sub-leading corrections.  Work along these lines is in progress.  This will allow us to extend the evolution to lower temperatures where $E/T$ is larger.  Physics-wise this will also allow us to go beyond the recoilless limit.  

\section*{Acknowledgments}
I thank my collaborators on Refs.~\cite{Brambilla:2021wkt,Omar:2021kra}.
M.S. has been supported by the U.S. Department of Energy, Office of Science, Office of Nuclear Physics Award No.~DE-SC0013470.
M.S. also thanks the Ohio Supercomputer Center for support under the auspices of Project No.~PGS0253.  

\end{multicols}
\medline
\begin{multicols}{2}
\bibliographystyle{JHEP-2}
\bibliography{strickland} 
\end{multicols}
\end{document}